\documentclass[twocolumn,prl,aps,showpacs,floatfix,superscriptaddress,preprintnumbers,amsmath,amssymb]{revtex4}

\usepackage{graphicx}

\begin{document}

\newcommand{\mfs}{Mn$_{1-x}$Fe$_{x}$Si}
\newcommand{\mcs}{Mn$_{1-x}$Co$_{x}$Si}
\newcommand{\fcs}{Fe$_{1-x}$Co$_{x}$Si}
\renewcommand{\vec}[1]{{\bf #1}}

\title{Long-range crystalline nature of the skyrmion lattice in MnSi}

\author{T. Adams}
\affiliation{Technische Universit\"at M\"unchen, Physik-Department E21, D-85748 Garching, Germany}

\author{S. M\"uhlbauer}
\affiliation{Technische Universit\"at M\"unchen, Physik-Department E21, D-85748 Garching, Germany}
\affiliation{Technische Universit\"at M\"unchen, Forschungsneutronenquelle Heinz Maier-Leibnitz, D-85748 Garching, Germany}
\affiliation{ETH Z\"urich, Institut f\"ur Festk\"orperphysik, Z\"urich, Switzerland}

\author{C. Pfleiderer}
\affiliation{Technische Universit\"at M\"unchen, Physik-Department E21, D-85748 Garching, Germany}

\author{F. Jonietz}
\affiliation{Technische Universit\"at M\"unchen, Physik-Department E21, D-85748 Garching, Germany}

\author{A. Bauer}
\affiliation{Technische Universit\"at M\"unchen, Physik-Department E21, D-85748 Garching, Germany}

\author{A. Neubauer}
\affiliation{Technische Universit\"at M\"unchen, Physik-Department E21, D-85748 Garching, Germany}

\author{R. Georgii}
\affiliation{Technische Universit\"at M\"unchen, Physik-Department E21, D-85748 Garching, Germany}
\affiliation{Technische Universit\"at M\"unchen, Forschungsneutronenquelle Heinz Maier-Leibnitz, D-85748 Garching, Germany}

\author{P. B\"oni}
\affiliation{Technische Universit\"at M\"unchen, Physik-Department E21, D-85748 Garching, Germany}

\author{U. Keiderling}
\affiliation{Helmholtz Zentrum Berlin, BENSC, D-14109 Berlin, Germany}

\author{K. Everschor}
\affiliation{Institute of Theoretical Physics, Universit\"at zu K\"oln, D-50937 K\"oln, Germany}
\author{M. Garst}
\affiliation{Institute of Theoretical Physics, Universit\"at zu K\"oln, D-50937 K\"oln, Germany}
\author{A. Rosch}
\affiliation{Institute of Theoretical Physics, Universit\"at zu K\"oln, D-50937 K\"oln, Germany}

\date{\today}

\begin{abstract}
  We report small angle neutron scattering of the skyrmion lattice in
  MnSi using an experimental set-up that minimizes the effects of
  demagnetizing fields and double scattering. Under these conditions
  the skyrmion lattice displays resolution-limited Gaussian rocking
  scans that correspond to a magnetic correlation length in excess of
  several hundred ${\rm \mu m}$. 
  This is consistent with exceptionally well-defined
  long-range order. We further establish the existence of higher-order
  scattering, discriminating parasitic double-scattering with
  Renninger scans.  The field and temperature
  dependence of the higher-order scattering arises from an
  interference effect. It is characteristic for the long-range
  crystalline nature of the skyrmion lattice as shown by
  simple mean field calculations.
\end{abstract}


\vskip2pc

\maketitle
Single-crystal small angle neutron scattering (SANS) in the A-phase of
the itinerant helimagnet MnSi recently established a highly unusual
six-fold symmetry of the scattering pattern perpendicular to the
 magnetic field $\bf B$
\cite{muehlbauer:09b}. Regardless of the orientation of the crystal
lattice with respect to the magnetic field the same six-fold
diffraction pattern was seen, characteristic of a magnetic structure
which is almost completely
decoupled from the underlying atomic lattice.  A theoretical
calculation in turn identified the A-phase in MnSi as a skyrmion
lattice stabilized by thermal fluctuations, i.e., a new form of
magnetic order composed of topologically stable knots in
the spin structure. The associated non-trivial topology was confirmed
by means of the topological Hall signal \cite{neubauer:09}.  Both the theoretical analysis \cite{muehlbauer:09b} 
and small angle neutron scattering in
{\mfs}, {\mcs} and the strongly doped semiconductor {\fcs} further
suggested  that the skyrmion lattice is a general
phenomenon \cite{Muenzer:09}, with spin torque effects at
ultra-low current densities as the most spectacular result
\cite{jonietz:10}.

These studies inspired Lorentz force
microscopy providing direct evidence of skyrmion
lattices and individual skyrmions in {\fcs}
\cite{Yu:10}, FeGe \cite{Yu:11} and MnSi \cite{Tokura}. In the thin
samples and for the perpendicular magnetic fields studied, 
the skyrmion lattice is thereby more stable \cite{Yu:10} than in the bulk,
but with increasing sample thickness the magnetic phase diagram was
found to approach that of bulk samples. This suggests that 
the same skyrmion lattice is realized in bulk samples and thin films. 

Taken together a pressing question regarding the existence of skyrmion
lattices in bulk materials concerns direct microscopic evidence of
their long-range crystalline nature, i.e., the precise spatial
variation of the magnetization on very long scales. The quality of
the long-range order strongly influences  both the pinning forces
\cite{jonietz:10,Zang:11} and the rotational spin
torque forces observed in the presence of currents~\cite{jonietz:10}.  

In principle, quantitative information on the magnetic structure can be
obtained from a reconstruction
from higher-order peaks in neutron scattering.  However, previous
neutron studies \cite{muehlbauer:09b} were subject to strong double
scattering, i.e., neutrons scattering twice from the magnetic
structure, which are difficult to discern from higher-order peaks.
Moreover, an exponential (rather than Gaussian) intensity
variation of the rocking scans and streaks of intensity emanating
radially from the first order peaks \cite{muehlbauer:09b} seemed to hint at 
unusual aspects of the morphology of the spin structure.

In this Letter we report a high-resolution small angle neutron
scattering study of pure MnSi to resolve these issues. 
As the skyrmion lattice is extremely weakly coupled to the atomic lattice and follows closely the applied magnetic field, it is essential to guarantee a homogeneous magnetic field inside the sample by reducing all effects of demagnetization fields. We used therefore a thin sample with a thickness of 
$\sim1\,{\rm mm}$ which was illuminated in a small central section only. 
This also allowed us to reduce the amount of double scattering.
Taking these
precautions we find sharp resolution-limited Gaussian rocking scans,
while all peculiar features such as streaks of intensity
vanished. Our results highlight, that
extreme care has to be exercised before claiming an 
unusual morphology of the skyrmion lattice in 
any B20 compound.

As our main result we unambiguously establish higher-order scattering,
by discriminating parasitic doubly scattering in Renninger scans.  The
magnetic field $B$ and temperature $T$ dependence of the higher-order
scattering may thereby be quantitatively explained with an interference
effect, providing unambiguous microscopic evidence of the long-range 
crystalline nature of the skyrmion lattice in bulk samples of MnSi.

For our studies two MnSi platelets denoted as sample A and B were cut
from the same ingot used previously \cite{Janoschek:10,Roessli:02,Semadeni:99,muehlbauer:09b,
jonietz:10}. Sample A was $\sim 14 \times 9 \times 
{\rm 1.4\,mm^3}$, and sample B was $\sim 12 \times 7 
\times {\rm 1\,mm^3}$, both with a crystalline $\langle
110\rangle $ direction normal to the platelet. The crystalline mosaic
spread was measured to be very small, $\sim 0.15^{\circ}$. The
specific heat and resistivity of test pieces from the same ingot are
in excellent agreement with literature with a good residual
resistivity ratio, ${\rm RRR}\sim$100. The samples were prealigned by
x-ray Laue backscattering. For precise alignment in the SANS studies at low $T$ we use that the magnetic structure at zero field is described by helices 
in $\langle 111 \rangle$ directions.

\begin{figure}
\includegraphics[width=0.45\textwidth]{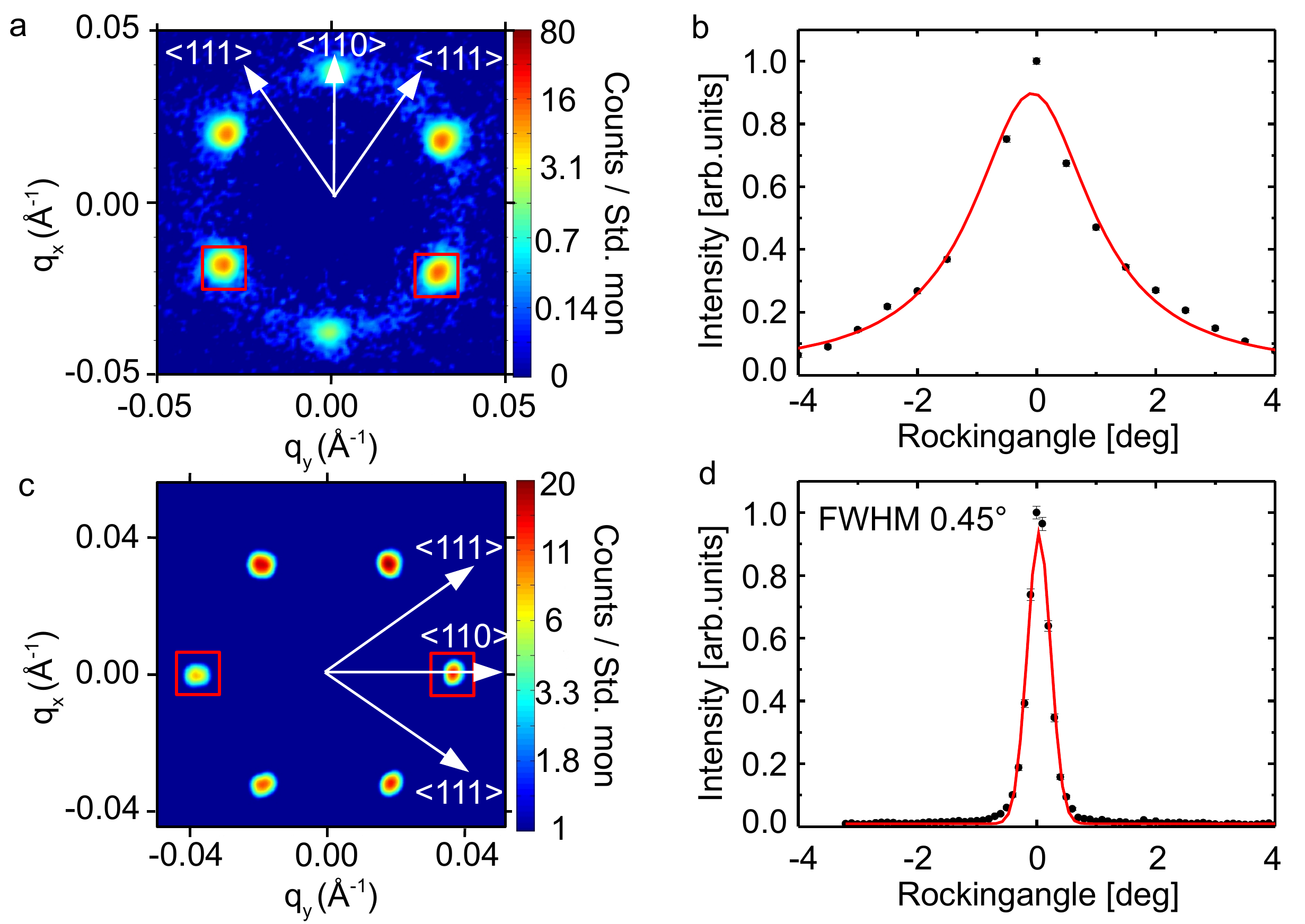}
\caption{Typical SANS data in the A-phase of MnSi. (a) Intensity pattern in a thick cylindrical sample  at $\mu_0H=0.16\,{\rm T}$ as reported in \cite{muehlbauer:09b}. (b) Rocking scan of the data shown in panel (a) as reported in \cite{muehlbauer:09b}. (c) Intensity pattern in thin platelet (sample A) at $\mu_0H=0.2\,{\rm T}$. (d) Rocking scan of the data shown in panel (c); note the narrow Gaussian dependence.}
\label{Fig_3}
\end{figure}


Our studies were carried out on the cold diffractometer MIRA at FRM II
(Munich) and the SANS instrument V4 at BENSC (Berlin).  Neutrons with
a wavelength $\lambda=9.7\,{\rm \AA}\pm5\%$ and $\lambda=4.5\,{\rm
  \AA}\pm10\%$ were used at MIRA and V4, respectively. The instrumental
resolution at MIRA was $\Delta\beta_{az}=4^{\circ}$ in azimuthal
direction, $\Delta\beta_{\bf q} =0.004\,{\rm \AA}^{-1}$ in radial
$|{\bf q}|$ direction and $\Delta\beta_{\bf k_f}=0.35^{\circ}$
perpendicular to $|{\bf q}|$ in the direction of ${\bf k_f}$. This
compares with a resolution of $\Delta\beta_{az}=4.9^{\circ}$,
$\Delta\beta_{\bf q} =0.003$\,\AA$^{-1}$, and $\Delta\beta_{\bf
  k_f}=0.21^{\circ}$ at V4. 
Temperatures are given with respect to the helimagnetic transition temperature $T_c$ determined by SANS.

The importance of the sample thickness for SANS studies  
is evident from previous work in the helical state, 
where distinct double scattering was observed, 
e.g., Fig.\,2\,(A) \& (D) in Ref.\,\cite{muehlbauer:09b}. 
Consistent with the entire literature on SANS 
in the zero field helical state
(cf. Ref.\,\cite{Lebech:95,Pfleiderer:07a} and references therein), 
typical data in the helical state of sample A (not shown) 
display a Gaussian rocking dependence 
with a rocking width of $\eta_m=1.6^{\circ}$
corresponding a correlation length $\sim10^4\,{\rm \AA}$. 
However, due to its reduced thickness 
double scattering is strongly reduced in sample A. 

The effects of the sample shape are most dramatic for  
for SANS studies of the A phase as illustrated in Fig.\,\ref{Fig_3}\,(a \& b), which reproduces Fig.\,2\,(E) of Ref.\,\cite{muehlbauer:09b}.
These previous results suggested a rocking width 
similar to the helical state, 
with an unusual non-Gaussian line shape (Fig.\,\ref{Fig_3}\,(b)).
For another sample studied in Ref.\,\cite{muehlbauer:09b}
strong double scattering and even unexplained streaks of intensity 
emanating radially from the first order scattering were seen.
In stark contrast, we do not find any such peculiar features 
for our thin platelets (see Fig.\,\ref{Fig_3}\,(c \& d)).
Moreover, the rocking dependence is now Gaussian 
with an extremely narrow width, $\eta_A=0.45^{\circ}$, slightly 
larger than the resolution limit $\Delta \beta_{{\bf k}f}=0.35^{\circ}$. 
Thus, the intrinsic magnetic correlation length of the skyrmion lattice
exceeds $100\,{\rm \mu m}$ and is therefore more than a factor of 
100 larger than for the helical state. In contrast, 
previous studies reflected the variation of the internal 
magnetic field directions due to demagnetizing effects.

\begin{figure}
\includegraphics[width=0.45\textwidth]{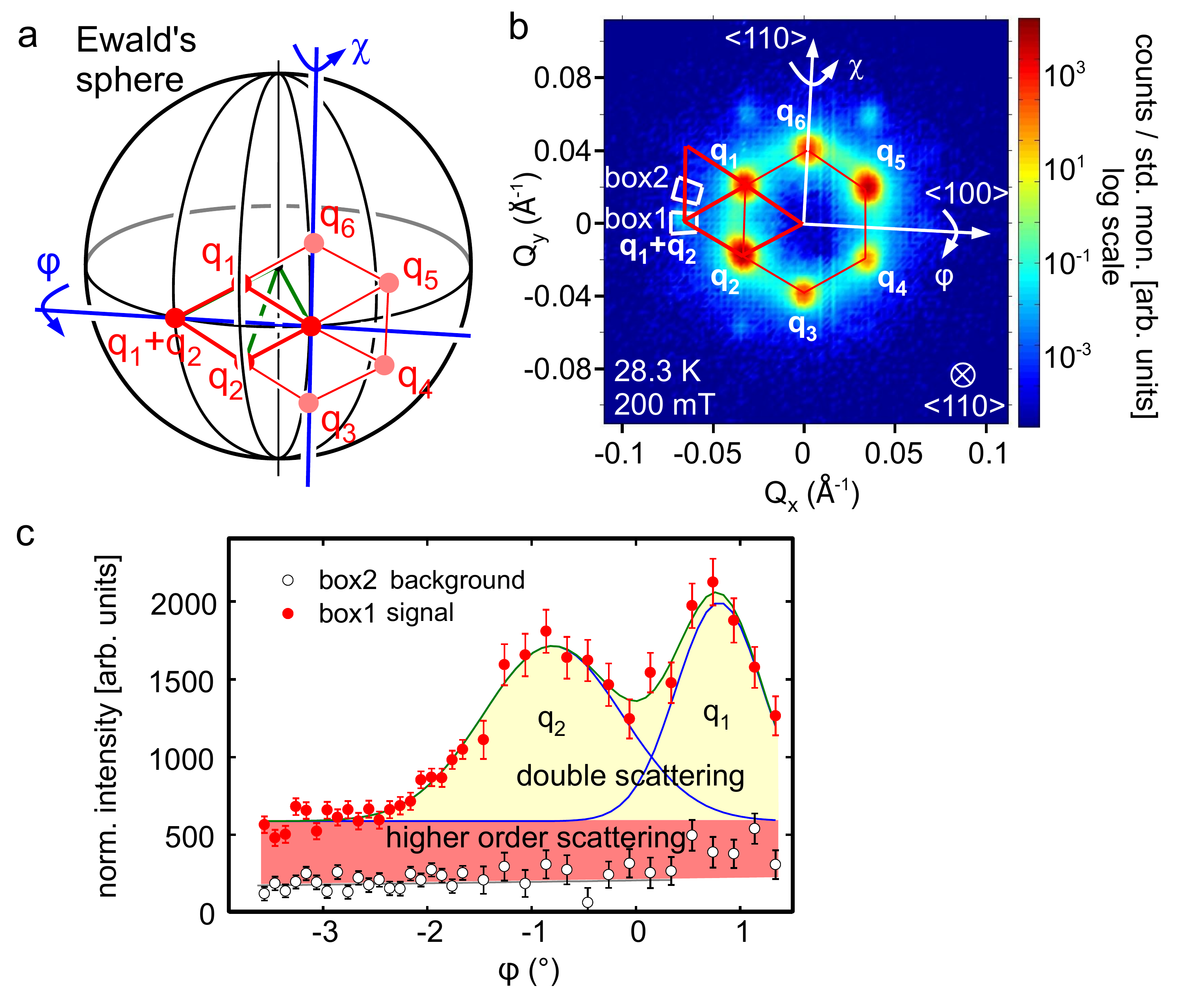}
\caption{
Operating principle and typical data of
Renninger scans. (a) Ewald-sphere depiction of Renninger scans, see text for details. (b) Typical scattering pattern obtained by a sum over a rocking scan
around $\phi$ after background subtraction recorded at high $T$. 
(c) Intensity as a function of rocking angle $\phi$ in a Renninger scan. 
The intensity was integrated over the areas
indicated by box 1 and box 2 in panel (b).}
\label{Fig_2}
\end{figure}

\begin{figure}
\includegraphics[width=0.45\textwidth]{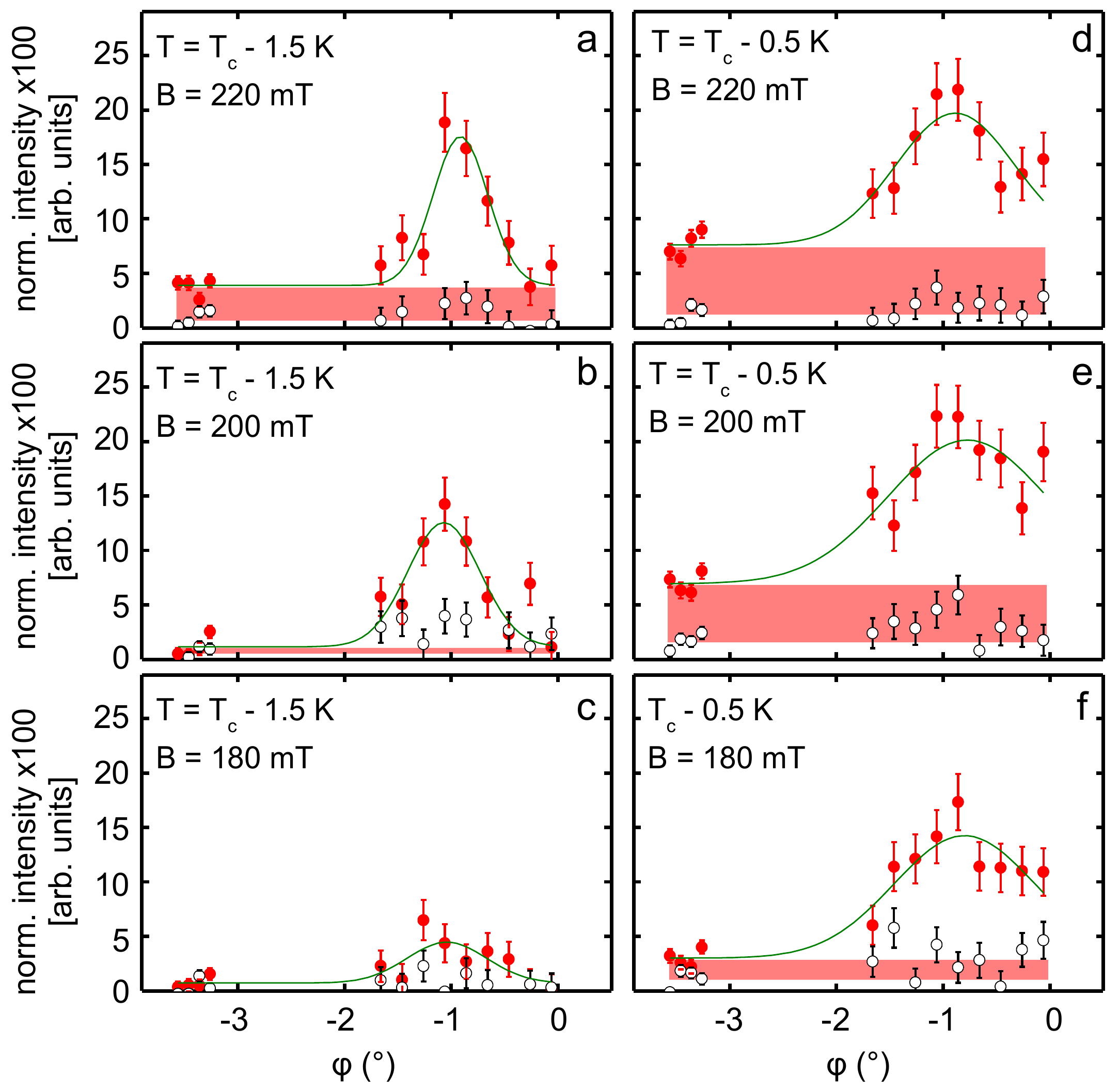}
\caption{Typical Renninger scans at various $T$ and $B$ in the A-phase of MnSi. Panels (a) through (c) were recorded at $T=T_c-1.5\,{\rm K}$ and magnetic fields $\mu_0H=180\,{\rm mT}$, $200\,{\rm mT}$ and $220\,{\rm mT}$, respectively. Panels (d) through (f) show Renninger scans for $T=T_c-0.5\,{\rm K}$ and magnetic fields of $\mu_0H=180\,{\rm mT}$, $200\,{\rm mT}$ and $220\,{\rm mT}$. True higher-order scattering is shaded in red and remains constant for large $\phi$.}
\label{Fig_4}
\end{figure}

To distinguish double scattering from higher-order scattering we used
so-called Renninger scans depicted in Fig.\,\,\ref{Fig_2}(a) \cite{Shirane:02}. 
The sample is thereby first rotated together with the magnetic field around the
vertical axis through an angle $\chi$ until the sum of two scattering
vectors ${\bf q_1} +{\bf q_2}$ touches the Ewald sphere thus
satisfying the scattering condition. This is followed by the actual
Renninger scan, which is a rocking scan with respect to ${\bf q_1}
+{\bf q_2}$ through the angle $\phi$, while recording the intensity
at ${\bf q_1} +{\bf q_2}$. This way double scattering is ``rocked out''
of the scattering condition, while higher-order scattering continues
to satisfy the scattering condition for all~$\phi$.
For the Renninger scans sample B was mounted with its crystalline
$\langle 110\rangle$ direction parallel to the incoming neutron
beam. A crystalline $\langle 100\rangle$ direction was oriented
approximately horizontally.  The background was determined for $T$ well above $T_c$ for each rocking angle and subsequently
subtracted. The intensity at ${\bf q_1} +{\bf q_2}$ as
indicated by box 1 in Fig.\,\ref{Fig_2} panel (b) was then compared
with the intensity in a box of equal size at a position
slightly to the side of ${\bf q_1} +{\bf q_2}$, labelled box
2. Typical variations of the intensities in box 1 and box 2 with the angle
$\phi$ are shown in Fig.\,\ref{Fig_2}, panel (c) for $T=T_c-0.5\,{\rm K}$
and $\mu_0H=200\,{\rm mT}$. The intensity observed at ${\bf q_1} +{\bf
q_2}$ clearly displays two contributions: (a) two Gaussian peaks due
to double scattering when either ${\bf q_1}$ or ${\bf q_2}$ 
intersect the Ewald sphere, and (b) a constant intensity arising
due to true higher-order reflections (red shading).

\begin{figure}
\includegraphics[width=0.45\textwidth]{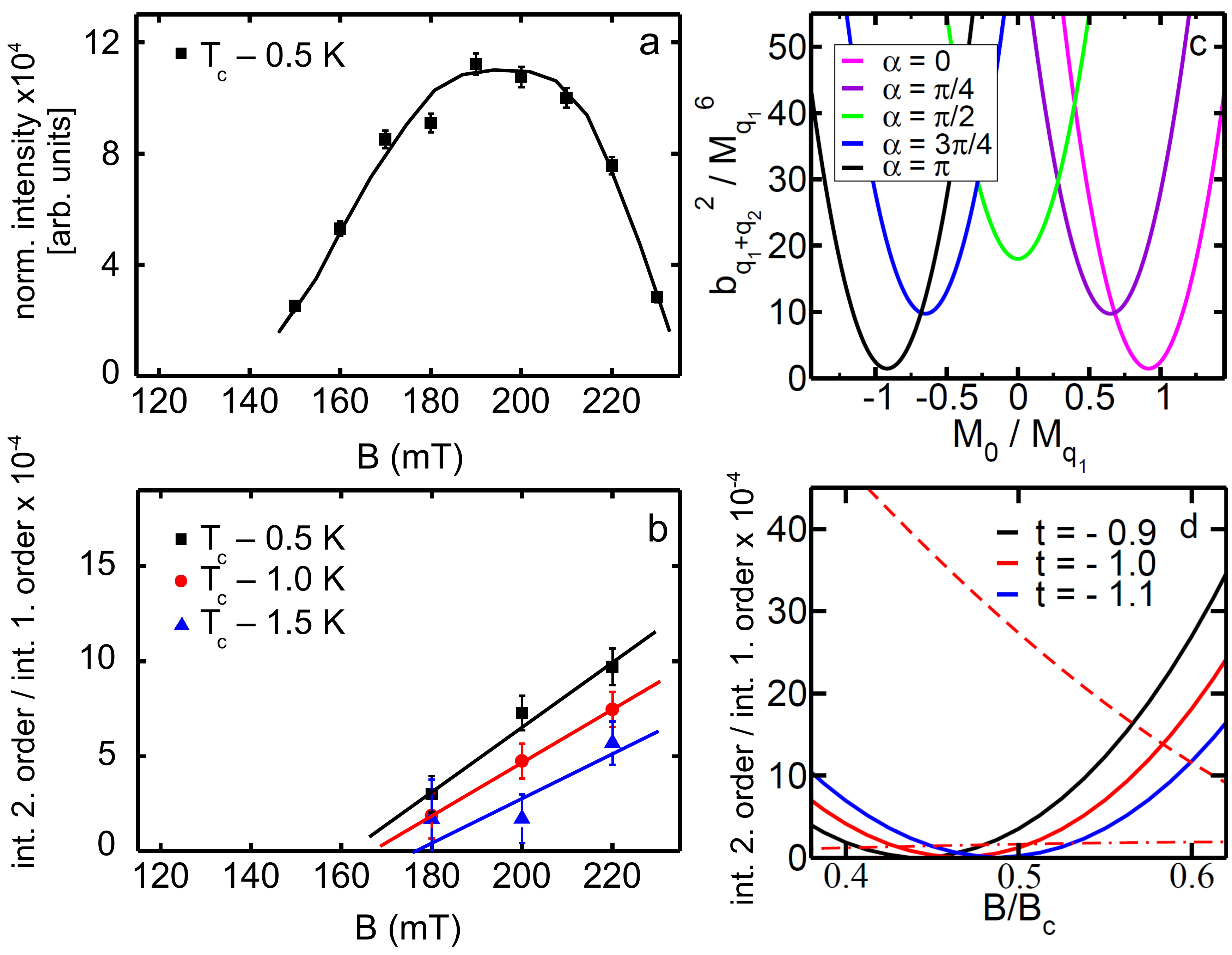}
\caption{(a) Intensity of first order peaks in the A-phase 
as a function of $B$ for $T=T_c-0.5\,{\rm K}$. (b) Ratio of the intensity at 
${\bf q_1} +{\bf q_2}$ and first order diffraction at various temperatures. 
(c) Calculated staggered field at $|\vec b_{q_1+\vec q_2}|^2$, 
Eq. (\ref{eqB}), which induces the scattering at $\vec q_1+\vec q_2$ for different values of the phase $\alpha$.  
(d) Calculated ratio of the higher-order diffraction at 
${\bf q_1}+{\bf q_2}$ and first order diffraction, 
$|M_{\vec q_1+\vec q_2}|^2/|M_{\vec q_1}|^2$ as a function of the magnetic field in mean field theory for $t=-.9, -1$ and $-1.1$ 
in the units of Eq.~(\ref{F}). 
Dashed (dot-dashed) line: 
$|M_{2 \vec q_1}|^2/|M_{\vec q_1}|^2$ 
($|M_{2 \vec q_1+\vec q_2}|^2/|M_{\vec q_1}|^2$) for $t=-1$.}
\label{Fig_5}
\end{figure}

Typical Renninger scans for $T=T_c-0.5\,{\rm K}$, $T_c-1.0\,{\rm K}$, 
and $T_c-1.5\,{\rm K}$ under magnetic fields of
$\mu_0H=180\,{\rm mT}$, $200\,{\rm mT}$ and $220\,{\rm
  mT}$ are shown in Fig.\,\,\ref{Fig_4}. With increasing $T$
and $B$ the higher-order scattering increases.
The $T$ and $B$ dependence of the scattering intensities 
are summarized in Fig.\,\ref{Fig_5}. 
Panel (a) shows the field dependence of the first order 
peaks at ${\bf q_1}$ and $T=T_c-0.5\,{\rm K}$. 
The peak height is shown, which was confirmed to be 
proportional to the integrated intensity. 
Compared with the field dependence for 
$B\parallel \langle111\rangle$ reported 
in Ref.\,\cite{grigoriev:06} for $T_c-0.2\,{\rm K}$, 
we find a more gradual variation of the intensity.
Panel (b) shows the ratio of the higher order peak 
intensity at ${\bf q_1} +{\bf q_2}$ with respect to the first
order intensity at ${\bf q_1}$ for $T=T_c-1.5\,{\rm K}$, $T_c-1.0\,{\rm
 K}$ and $T_c-0.5\,{\rm K}$. Above $\mu_0H\approx180\,{\rm mT}$ this
ratio increases as a function of $T$ and $B$.  Unfortunately it was not possible to track the higher-order scattering below
$\mu_0H=180\,{\rm mT}$, since the first order intensity was too weak.

In summary our main experimental results are: (i) A strong magnetic
field dependence of the second order intensity, apparently vanishing
for a certain field $B\approx B_{\rm int}$ inside the A-phase. 
(ii) An increase of the second
order intensity with increasing $T$. This may appear
counter-intuitive, since the non-linear effects leading to higher
order peaks should be less pronounced when all amplitudes decrease
with increasing $T$. And finally, (iii) a tiny weight of the
higher-order peaks of the order of $10^{-3}$.

For a theoretical analysis, we first discuss what can be inferred from
the experiments without a detailed calculation and compare in a second
step our experimental results to mean field theory. Almost all of the
scattering intensity arises from six resolution-limited main
scattering peaks, and therefore the magnetic structure is well 
approximated by a superposition of three helices and the uniform magnetization.

A single pair of peaks at $\pm \bf q_i$ describes a spin helix of a given
chirality \cite{tanaka:JPSJ85} determined by
the chirality of the atomic structure. A representative of such a
helix is $\frac{\Phi}{\sqrt{2}} (0, \sin q x, -\cos q x)$ 
or, in Fourier space, 
$\vec M_{\vec q_1}=\frac{\Phi}{\sqrt{2}} (0,-i,-1)$ 
for $\vec q_1=(q,0,0)$, where $\Phi^2$ is the weight of the peak. 
Our observation of higher-order scattering proves 
a crystalline character of the magnetic state, 
which may therefore be described by a linear superposition of
three such helices rotated by $n 2 \pi/3$ ($n=0,1,2$) around the $z$
axis. The phase relationship between the helices thereby distinguishes
the skyrmion lattice from other forms of magnetic order. Two of the
three phases just describe translations of the magnetic structure in
the $xy$ plane. However, the third phase $\alpha$, defined by
multiplying $\vec M_{\vec q_1}$ by $e^{i \alpha}$, while keeping the
other two helices unmodified, strongly affects the magnetic structure
(in our convention $\alpha=0$ describes the skyrmion lattice
\cite{muehlbauer:09b} where in the center of the skyrmion all three
helices point antiparallel to the external magnetic field).

Unfortunately, no information on the relative phase of the three
helices can be inferred from first order scattering, since the signal
is only sensitive to $|\vec M_{\vec q_i}|^2$. In contrast, the
higher-order peaks are very sensitive to the phase $\alpha$ because
they are subject to interference effects.  When considering the
Fourier transformation of the leading magnetic interaction term 
$\vec M(r)^4$, collecting all terms {\em linear} in 
$\vec{M}_{\vec q_1+\vec{q_2}}$, 
$\int {\vec M}^4 {\rm d}^3\vec r = 
\vec{M}_{\vec q_1+\vec{q_2}} \vec b_{\vec q_1+\vec{q_2}} +\dots$, 
one obtains an oscillating effective field
$\vec b_{\vec q_1+\vec{q_2}}$ arising from $\vec M_{\vec q_{1...6}}$ and the uniform magnetization $\vec M_0$.  
From $\vec q_1 + \vec q_2 = 2 \vec q_1 +
\vec q_3= 2 \vec q_2+ \vec q_6$ (see Fig.~\ref{Fig_2}b) one obtains
that the interference of several processes determines the strength of
$\vec b_{\vec q_1+\vec{q_2}}$. Adding all of these terms (examples are
$\vec M_0 (\vec M_{-\vec q_1}\vec M_{-\vec q_2})$ or $\vec
M_{-\vec{q}_1} (\vec M_{-\vec{q}_1}\vec M_{\vec q_3})$) we obtain
\begin{align}
 \frac{\vert\vec b_{\vec
    q_1+\vec{q_2}}\vert^2}{2 \Phi^6} = 9 + 
74 \frac{ M_0^2}{\Phi^2} - 96 \sqrt{2} \frac{M_0}{\Phi} \cos
\alpha + 54 \cos^2  \alpha \label{eqB}
\end{align}  
A plot of this function for various values of $\alpha$ 
is shown in Fig.~\ref{Fig_5}\,(c). As negative
$M_0$ are unphysical, one finds only for values of $\alpha$ close to
$\alpha=0$ that $\vec b_{\vec q_1+\vec{q_2}}$ becomes very small for a
certain magnetic field where $M_0/\Phi=\sqrt{2}\,24/37\approx 0.92$
(to be compared to $0.94$ and $0.96$ obtained from the mean field
theory given below for $t=-1$ and $t=-5$, respectively). The observed strong
suppression of $\vec M_{\vec q_1+\vec{q_2}}$ is therefore the
signature of a very small $\alpha$ characteristic for the skyrmion lattice
\cite{muehlbauer:09b}.

A complete quantitative theory of the skyrmion phase requires 
account of the effects of thermal fluctuations \cite{muehlbauer:09b}. Yet, a
semi-quantitative explanation of our experimental observations may 
already be obtained on the level of a simple mean-field 
 approximation. After a 
rescaling of coordinates, magnetization, magnetic field
and free energy, the Ginzburg Landau free energy density in the presence of
the Dzyaloshinskii-Moriya interaction $\sim \vec M(\nabla\!\times\! \vec
M)$ is given by  \cite{Nakanishi:80,Bak:Jul80,muehlbauer:09b}
\begin{equation}
F= (1+t) \vec M^2+(\nabla \vec M)^2+2 \vec M(\nabla\!\times\! \vec
M)+\vec M^4-
\bf B M \label{F}
\end{equation}
where $t$ measures the distance to the $B=0$ mean-field critical
temperature.  Other contributions can be neglected as they are higher
order in spin-orbit coupling or give only small contributions close to
the critical temperature where the skyrmion lattice is stable.
Minimizing $F$ with the Ansatz 
$\vec{M}(\vec x)=\sum_{n,m} 
e^{i ( n \vec q_1+ m \vec{q}_2)\vec x}\, 
\vec{M}_{n \vec q_1+ m \vec{q}_2}$
for integer $n$ and $m$ provides the relative weight of $|M_{\vec
  q_1+\vec q_2}|^ 2/|M_{\vec q_1}|^ 2$ and other higher-order peaks,
cf Fig.\,\ref{Fig_5}(d).  
An open issue for future studies is the gradual variation 
of the amplitude $M_{\bf q_1}$ for small $B$, 
cf Fig.\,\ref{Fig_5}\,a, 
where theory predicts 
a sharp first-order transition \cite{muehlbauer:09b}.

Our calculation reproduces the main
experimental observations, including the tiny weight of the
higher-order peaks and the
approximate vanishing of the signal for a certain magnetic field,
$B\approx B_{\rm int}$ due to the above 
discussed interference effect.
Most importantly, it explains the unexpected rise of the signal for
increasing $T$:
the overall drop of the higher-order scattering for larger $T$ is
overcompensated by a shift of $B_{\rm int}$ towards smaller values for
increasing $T$. Note that both experimental and theoretical
uncertainities (precise absolute values of weights and 
fluctuation corrections, respectively) presently prohibit a
precise quantitative determination of, e.g., the $T$
dependence of the parameter~$t$.

We wish to thank E. M. Forgan, P. Granz, 
M.~Janoschek, H. Kolb, M. Laver, 
J. Peters, B. Russ, R. Schwikowski,  
D. Wallacher and A. Zheludev
for support and stimulating discussions. 
TA and AB acknowledge support through the TUM Graduate School. 
Financial support through DFG
TRR80, SFB608 and FOR960 is gratefully acknowledged.

\end{document}